\documentclass[showpacs,preprint,amsmath,PRB]{revtex4}

\begin{document}

\title{Probing onset of strong localization and electron-electron interactions
with the presence of direct insulator-quantum Hall transition}

\author{Shun-Tsung Lo$^1$, Kuang Yao Chen$^1$, T. L. Lin$^1$, Li-Hung Lin$^2$,
Dong-Sheng Luo$^3$, Y. Ochiai$^4$, N. Aoki$^4$, Yi-Ting Wang$^1$, Zai Fong Peng$^2$, Yiping Lin$^3$, J. C. Chen$^3$, 
Sheng-Di Lin$^5$,  C. F. Huang$^6$, and C.-T.~Liang$^{1, \ast}$}
 
\address{$^1$Department of Physics, National Taiwan University, Taipei 106, Taiwan}
\address{$^2$Department of Applied Physics, National Chiayi University, Chiayi 600, Taiwan}
\address{$^3$Department of Physics, National Tsinghwa University, Hsinchu 300, Taiwan}
\address{$^4$Graduate School of Advanced Integration Science, Chiba University, 1-33 Yayoi-cho, Inage-ku, 
Chiba 263-8522, Japan}
\address{$^5$Department of Electronics, National Chiao Tung University, Hsinchu~300, Taiwan}
\address{$^6$National Measurement Laboratory, Centre for Measurement Standards, Industrial
Technology Research Institute, Hsinchu 300, Taiwan}

\begin{abstract}
We have performed low-temperature transport measurements on a disordered two-dimensional electron system (2DES). 
Features of the strong localization leading to the quantum Hall effect are observed after the 2DES undergoes a 
direct insulator-quantum Hall transition with increasing the perpendicular magnetic field. However, such a 
transition does not correspond to the onset of strong localization. The temperature dependences of the Hall resistivity and Hall 
conductivity reveal the importance of the electron-electron interaction effects to the observed transition in our study. 

$^\ast$ E-mail addresses: ctliang@phys.ntu.edu.tw
\end{abstract}

\maketitle

\section{Introduction}
When a strong magnetic field $B$ is applied perpendicular to the plane of a two-dimensional electron system (2DES), Landau quantization may cause 
the formation of Landau bands. It is now well established that Landau quantization can modify the electrical properties of a 2D system. 
With increasing $B$, usually Landau quantization may give rise to Shubnikov-de Haas (SdH) oscillations with the amplitude \cite{Coleridge,Coleridge2,fowler,ando,isihara} 
\begin{equation}
\bigtriangleup\rho_{xx}(B,T)=4\rho_{0}{\rm exp}(-\pi /\mu B)D(B,T)    
\end{equation} 
in the longitudinal resistivity $\rho _{xx}$
before the appearance of the integer quantum Hall effect (IQHE) \cite{Coleridge,Klitzing} at a low temperature $T$. Here $\rho _{0}$ is expected to be 
the longitudinal resistivity $\rho _{xx}$ at $B=0$ while there may exsit deviations \cite{jhchen}, $\mu$ is thequnatum mobility, and
$D(B,T)=2\pi^{2}k_Bm^{*}T/\hbar eB$sinh$(2\pi^{2}k_{B}m^{*}T/\hbar eB )$ with $m^{*}$, $k_B$, and 
$\hbar$ as the effective mass, Boltzmann constant, and reduced Plank constant. It is worth mentioning that the SdH theory is 
derived based on Landau quantization without considering the strong localization effects induced by the quantum interference. On the other hand, it is 
believed that both extended and localized states arising from such effects are key 
ingredients for describing the IQHE, in which the magnetic-field-induced transitions \cite{klz,Huo,jiang,hughes,Shahar} are good examples 
of quantum phase transitions. Such transitions occur as the Fermi energy passes through the extended states of 
Landau bands. In the global phase diagram (GPD) \cite{klz} of the quantum Hall effect, all the magnetic-field-induced 
transitions are regarded equivalent though they are divided into two types, plateau-plateau (P-P) transitions and insulator-quantum Hall (I-QH) transitions \cite{Shahar}. 

There has been much interest in the IQHE at low magnetic fields \cite{liu,sheng,song,huckestein,hang,hhc}. A thorough understanding of the low-field IQHE 
should provide important information regarding the I-QH transition \cite{klz,jiang,hughes,kim}. In particular, whether a direct transition from 
the insulating regime (denoted by symbol 0) to a $\nu \geq 3$ QH state can occur is an interesting, fundamental yet unsettled issue in 
the field of two-dimensional (2D) physics \cite{liu,sheng,song,lee,cfhuang,tyhuang,kychen}. Experimental and theoretical evidence for such a direct phase 
transition has been reported \cite{liu,sheng,song,lee,cfhuang,tyhuang}. On the other hand, it was argued by Huckestein \cite{huckestein}
that the observed direct I-QH transition is 
not a real quantum phase transition, but a crossover from weak localization to strong localization regime in which Landau quantization 
becomes dominant. Within Huckestein's model, the onset of strong localization which causes the formation of a QH 
state should correspond to the direct I-QH transition. We note that in such a model, both weak localization and electron-electron interactions 
are considered as correction terms to the classical Drude conductivities. It is well known that electron-electron (e-e) interactions could 
play an important role in the metal-insulator transition (MIT) \cite{dobro}. Moreover, in the seminal work of Dubi, Meir and Avishai \cite{dubi}, various transitions such 
as I-QH, MIT, and percolation transitions \cite{meir,shimshoni} can be explained within a unifying model. Therefore, it is interesting to probe electron-electron interaction effects with 
the presence of I-QH transition. Moreover, the effect of Landau quantization and onset of strong localization are fundamental issues regarding the direct I-QH 
transition. 

In this communication, we report magneto transport measurements on a disordered 2DES. With increasing $B$, the strength of the strong localization increases 
such that we can observe the well-developed QH state of $\nu=2$. However, the direct I-QH transition observed at $B \sim 2.29$ T is not due to the onset of the 
strong localization because SdH formula is valid as $B<4.76$ T. The $T$-dependences of the Hall resistivity $\rho _{xy}$ and Hall conductivity $\sigma _{xy}$ 
show the importance of e-e interactions to such a transition in our study. 

\section{Experimental details}
Sample LM4645, a delta-doped quantum well with additional modulation doping, is used in this study. The following layer sequence was grown on a 
semi-insulating GaAs (100) substrate: 500 nm GaAs, 80 nm Al$_{0.33}$Ga$_{0.67}$As, 5 nm GaAs, Si delta-doping with a concentration of 3 x 10$^{11}$ cm$^{-2}$, 
15 nm GaAs, 20 nm Al$_{0.33}$Ga$_{0.67}$As, 40 nm Si-doped Al$_{0.33}$Ga$_{0.67}$As 
with a doping concentration of 10$^{18}$~cm$^{-3}$, finally 10 nm GaAs cap layer. Experiments were performed in a top-loading He$^3$ 
cryostat equipped with a superconducting magnet. 
Four-terminal magneto resistivities were measured using standard ac phase-sensitive lock-in techniques. 
Magnetic field is applied perpendicular to the plane of the 2DES.

\section{Results and discussion}
Figure 1 shows longitudinal and Hall resistivities ($\rho_{xx}$ and  $\rho_{xy}$) as a function of magnetic field $B$ at various temperatures $T$. 
For $2.54$~T~$\leq B \leq 4.76$~T , magneto-oscillations in $\rho_{xx}$ are observed. In order to further study these oscillations, we plot their amplitudes 
as a function of $1/B$ as shown in Fig.~2. As shown in this figure, there is a good fit to Eq.~1, and thus these magneto-oscillations are ascribed to SdH oscillations. 
From the observed SdH oscillations, the carrier density of the 2DES
is measured to be $3.97 \times 10^{15}$~m$^{-2}$. According to the fit shown in Fig.~2, the quantum mobility is estimated to be $\approx 0.19$~m$^{2}$/Vs. 
Since the SdH theory is derived based on Landau quantization without considering strong localization effects, 
it is believed that high-field strong localization effects  leading to the IQHE are not  significant 
for $B \leq 4.76$~T. The resistance peak at around 6~T appears to move with increasing $B$. This movement cannot be described within 
the standard SdH theory and the measured amplitudes at $B=4.76$~T can be affected by this movement. 
Hence the data point at $B=4.76$~T in Fig. 2 are in open symbols whilst it can be fitted to Eq.~(1).

The Hall slope at low $B$ increases with decreasing $T$, and we can see from Fig. 1 that the curves of $\rho_{xy}$ at $T=0.33$~K and $T=1.242$~K do 
not collapse into a single curve as $B$ is smaller than 4 T. Such a change on the Hall slope does not result from a change in $n$ since $n$ determined from
the SdH oscillations is $T$-independent over the whole measurement range. As will be described later, such $T$-dependent $\rho_{xy}$ can 
be ascribed to electron-electron interactions.

As shown in the inset to figure 1, for $7.6$~T~$\leq B \leq 10.6$~T, we can see a 
well-quantized $\nu=2$ Hall plateau with corresponding vanishing resistivity. Therefore the strong localization effect which gives 
rise to the formation of the quantum Hall state should occur with increasing $B$. 

In order to further study the strong localization effect in our system, we follow the 
seminal work of Shahar \cite{Shahar} as described as follows. First, as shown in Fig.~3. we convert the measured $\rho_{xx}$ and  $\rho_{xy}$ into $\sigma_{xx}$ 
and  $\sigma_{xy}$ by matrix inversion
\begin{equation}
\sigma_{xx} = \frac{\rho_{xx}}{\rho_{xx}^2 +  \rho_{xy}^2},       
\end{equation}  
\begin{equation}
\sigma_{xy} = \frac{\rho_{xy}}{\rho_{xx}^2 +  \rho_{xy}^2}. 
\end{equation}  
Using the following equations, we then obtain the conductivity of the topmost Landau level by subtracting from the
conductivity data the contribution of the lowest, full Landau level. 
\begin{equation}
\sigma_{xx}^{t} = \sigma_{xx},           
\end{equation}  
\begin{equation}
\sigma_{xy}^{t} = \sigma_{xy} - \frac{2 e^2}{h}.           
\end{equation}
Finally we convert $\sigma_{xx}^{t}$ and $\sigma_{xy}^{t}$ into the corresponding resisitivities for the topmost Landau 
levels $\rho_{xx}^{t}$ and $\rho_{xy}^{t}$. Such results are shown in Fig. 4. 
We can clearly see a clear crossing point in $\rho_{xx}^{t}$ at around $5.2$~T, which is denoted by a vertical dotted line. 
Such a $T$-independent point can be ascribed to the formation of the extended states
under the existence of the localized states \cite{Shahar}. Since both extended and localized states are 
due to strong localization effects  leading to the IQHE, such effects should become significant as $B \geq 5.2$ T in our system. 
On the other hand, at $B<4.76$ T the validity of SdH formula reveals that the strength of strong localization is weak. Therefore, the onset 
of the strong localization occurs as $B=4.76$~T~$\sim 5.2$~T.  

It has been pointed out that the strong localization occurs at the magnetic field $B \sim1/\mu$, near which the localization length changes quickly \cite{huckestein}. 
Just as mentioned in the above, the mobility $\mu=0.19$~m$^2$/Vs and thus the onset of the strong localization is expected as $B \sim 5.3$~T. 
Such a magnetic field is close to the estimation based on SdH formula and the crossing point in $\rho_{xx}^{t}$ .
We note that the temperature-independent point in $\rho_{xx}^{t}$
is close to the resistance quantum $\frac{h}{2e^2}$ as expected for the topmost Landau level \cite{Shahar}.  In addition, as shown in Fig.~4, 
$\rho_{xy}^{t}$ does not deviate much from the expected value $\frac{h}{2e^2}$ at the lowest temperature $T=0.33$~K as $B < 5.2$~T.

We can see from Fig. 1 that the 2DES behaves as an insulator as $B<B_{c} \equiv 2.29 $~T in the sense that $\rho _{xx}$ increases 
with decreasing $T$. The longitudinal resistivity $\rho_{xx}$ is almost independent of $T$ at $B_{c}$. Since 
there is no QH state of the lowest integer 
filling factor 1 or  2 near $B_c$, the 2DES undergoes a direct I-QH transition at $B _{c}$ \cite{liu,song}. 
The filling factor $\nu$ is about 8 near $B_{c}$, so the observed transition is a 0-8 transition \cite{song,lee}. 
If such a transition is due to 
the onset of the strong localizationc \cite{huckestein}, $B_{c}$ should be within $B=4.76~T \sim 5.2$ T as mentioned above. 
In our study, however, $B_{c}$ is at a much lower magnetic field $B= 2.29 $~T. Therefore, the observed direct I-QH 
transition is not due to the onset of strong localization. 

It has been shown that by converting the measured $\rho_{xx}$ and  $\rho_{xy}$ into longitudinal and transverse conductivities  
$\sigma_{xx}$ and  $\sigma_{xy}$,  one can provide further information on the I-QH transition \cite{hughes,arapov}. Figure 3 
shows converted $\sigma_{xx}$ and $\sigma_{xy}$ as a function of $B$. We can see that $\sigma_{xy}$ is $T$-independent over 
a wide range of magnetic field ($0$~T$ \leq B \leq 2.8$~T), spanning from the insulating region to the QH-like regime.
On the other hand, as shown in the inset to Fig.~3, the Hall
slope $R_H$ shows an approximately ln$T$ dependence. The deviation from the linear fit through the full symbols can be ascribed to current heating. As
the current is decreased from 20~nA to 10~nA (full circle in blue), we are able to restore the ln$T$ dependence at low $T$ \cite{kychen}.
The observed ln$T$ dependence of $R_H$ does not result from a change in $n$ since $n$ determined from the SdH oscillations is
$T$-independent over the whole measurement range. Therefore, the observed $T$-indepndent $\sigma_{xy}$, together with the ln$T$-dependent $\rho_{xy}$
can be ascribed to electron-electron interaction, and we note that the corrections resulting from such an interaction have been 
discussed in the literature \cite{simmons}. Our experimental result therefore supports that the direct I-QH transition is not always due to 
the onset of strong localization when the e-e interaction is significant.

Interestingly, whilst there is a crossing field at $B_{c}$ in $\rho_{xx}$, there is no corresponding crossing point in $\sigma_{xx}$
\cite{tyhuang2010}. The reason for this is that $\rho_{xy}$ shows logarithmic dependence on $T$. Therefore according to Eq.~2, there is 
no corresponding crossing point in $\sigma_{xx}$ even there exists a $T$-independent point in $\rho_{xx}$. 

To further study the direct I-QH transition and onset of the strong localization leading to the IQHE, we have re-analyzed the data published 
in Ref. \cite{kychen}, 
where the studied sample is almost identical except a different delta-doping concentration of $5 \times 10^{11}$~cm$^{-2}$. 
There also exsits a crossing point in $\rho_{xx}^{t}$ ast $B \sim 1/ \mu$, near which the onset of the strong localization is expected. 
The direct I-QH transition, however, appears at a much lower magnetic field $B < 1/(2 \mu)$ and does not correspond to the onset of strong localization.  
Tilted-field measurements show that the sample studied in Ref. \cite{kychen} is two-dimensional such that the direct I-QH transition 
and features of Landau quantization only depend on the perpendicular component of the applied $B$. 

It has been reported in some case when $\rho_{xx}$ approaches zero and the strong localization effects may occur, the large resistance oscillations can be still 
well approximated by the conventional SdH formula \cite{hang,hhc}. In this case, rising background resistance \cite{hang} needs to be introduced 
while such background resistance does not occur in our system. In our system, the amplitudes of the resistance oscillations are a lot smaller than the 
non-oscillating background as $B \leq 4.76$~T, under which the resistance minima are much bigger than zero.

Based on the tight-binding model, Nita, Aldea and Zittartz \cite{nita} have predicted that resistance oscillations can cover the I-QH transition. 
There exists also experimental evidence for this prediction \cite{bohra}. We note that in this case, e-e interaction effects are not significant since $\rho_{xy}$
is nominally $T$-independent. It may be possible that the existence of e-e interactions may dictate the observation of SdH-like oscillations spanning
from the insulating regime to the QH-like regime. 

\section{Conclusion}
In conclusion, we have performed magneto transport measurements on a weakly-disordered 2DES. With increasing magnetic field, the 
2DES undergoes a direct 0-8 I-QH transition at a crossing field $B_{c}$. For $B > B_{c}$, magneto-oscillations governed by conventional 
Shubnikov-de Haas theory are observed. Since strong localization effect is not considered in the SdH theory, our results explicitly 
demonstrate that the direct I-QH transition does not correspond to the onset of strong localization. The observed nominally $T$-independent $\sigma_{xy}$ 
spanning from the insulating regime to the SdH regime, together with the observed logarithmic $T$-dependent Hall slope demonstrate that electron-electron interactions,
rather than the weak localization effects, are the dominant mechanism near the direct I-QH transition in our study.

Acknowledgment \\
This work was funded by the NSC, Taiwan. We would like to thank J.-Y. Wu and Y.-C. Su for experimental help.

Figure 1 Longitudinal resistivity $\rho_{xx}$ measurements as a function of magnetic field $B$ at various temperatures $T$. The Hall resistivity measurements at
the lowest and highest $T$ so as to highlight its weak $T$ dependence. The inset shows both $\rho_{xx}$ and $\rho_{xy}$ measurements  in the high-field 
regime at the lowest temperature $T=0.33$~K.

Figure 2  $\Delta \rho_{xx}/D(B, T)$ as a function of $1/B$ at various temperatures $T$ where $\Delta \rho_{xx}$ represents 
the amplitude of SdH oscillations. The solid curve corresponds to a fit to Eq.~(1)

Figure 3 Converted (a) $\sigma_{xx}(B)$ and (b) $\sigma_{xy}(B)$ at various temperatures $T$ ranging from $T=0.33$~K to 
$T=1.242$~K. Inset: semilogarithmic plot of Hall slope $R_H$ as a function of ln$T$. The linear fit to the full symbols is discussed in the text.

Figure 4 Converted  $\rho^{t}_{xx}$ as a function of $B$ at various temperatures $T$ ranging from $T=0.33$~K to 
$T=1.242$~K. $\rho_{xy}^{t}$ is at $T=0.33$~K. The vertical dotted line denotes the magnetic field where $\rho^{t}_{xx}$ is $T$-independent.

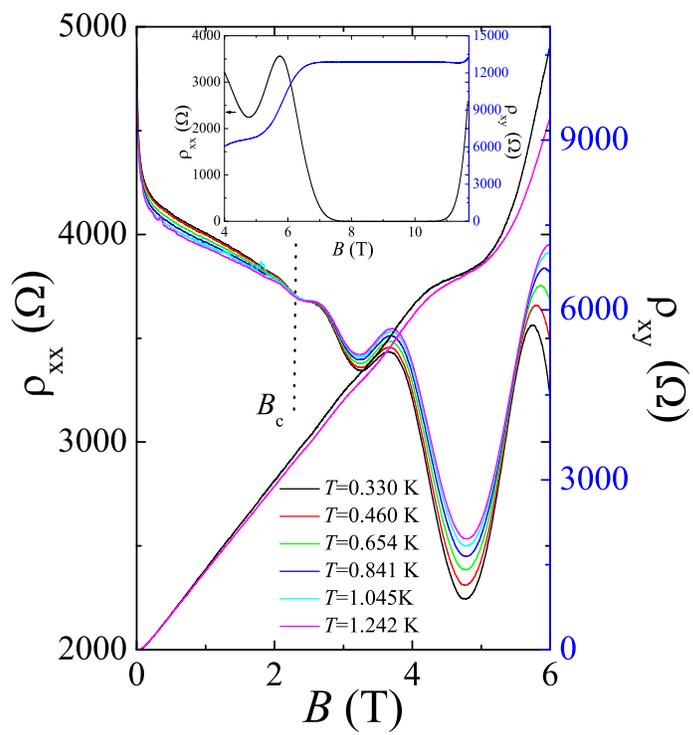
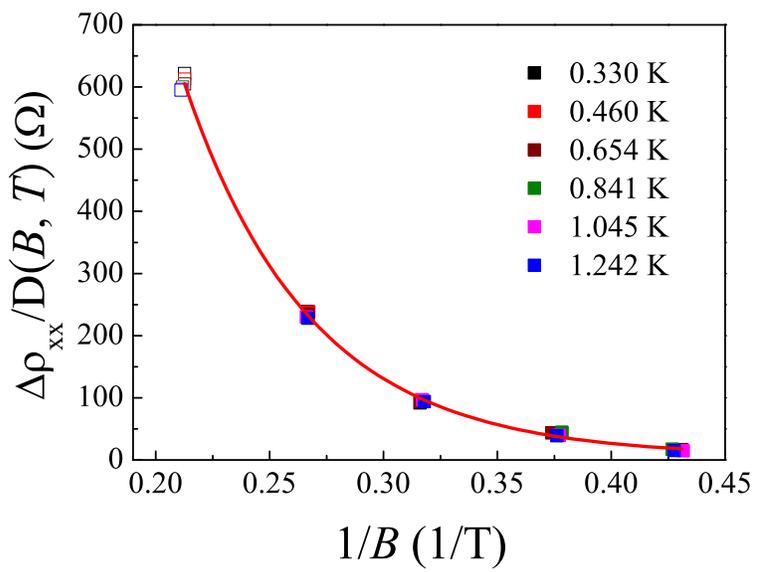
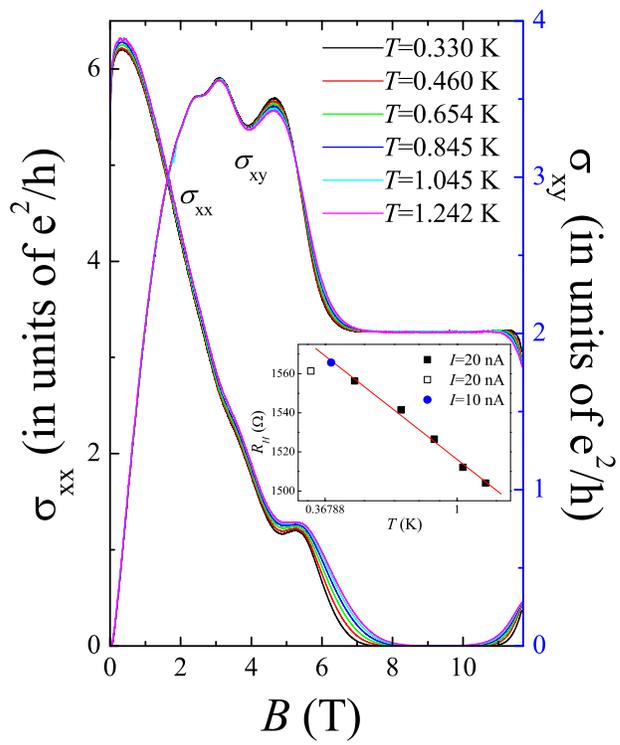
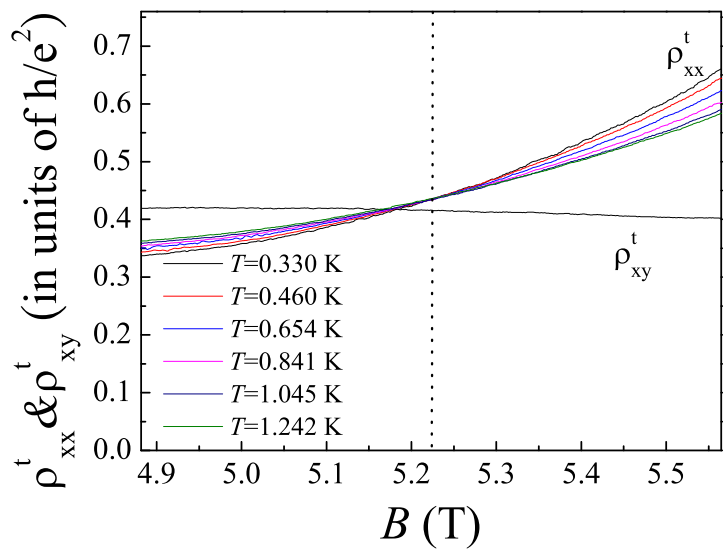
\end{document}